# *Evolutionary Optimization of AI-Collapsed Software Development Stacks: Labor Tipping Points and Workforce Realignment*

Matthew Hall Kilbane
*Athena's Zephyr LLC*
Fort Washington, MD, USA
ORCID ID: 0009-0000-9710-4866

***Abstract* – This paper presents a quantitative framework for optimizing human–AI workforce allocation in software development. I formalize baseline and AI-collapsed labor models, derive tipping-point equations for safe headcount reduction, and embed them in a multi-objective evolutionary-optimization setup. NSGA-II experiments reveal reproducible, phase-specific automation strategies that reduce cost while maintaining quality and stable workloads.**

## I. INTRODUCTION

The rapid emergence of agentic artificial intelligence is transforming how software systems are built and maintained, shifting organizations away from strictly layered development lifecycles toward tightly integrated, AI-orchestrated workflows.[1] In these "collapsed stacks," autonomous agents can participate in or fully execute activities that previously required multiple specialized human roles, including requirements analysis, coding, testing, deployment, and maintenance.[1] This transformation raises a fundamental quantitative question: given a target throughput and quality level, how should an organization allocate work between humans and AI, and at what point do AI-driven efficiencies justify discrete changes in team size and structure?[1]

Evolutionary computation (EC) offers a natural framework for addressing this question because the resulting design space mixes continuous decisions (fractions of work automated in each phase) with discrete ones (integer headcounts and role allocations), under competing objectives such as cost, cycle time, and quality.[2] Traditional engineering models treat automation as a static parameter, whereas in AI-intensive environments automation can be tuned per phase and evolves over time as capabilities improve.[^1] A principled approach requires both explicit analytic models of labor and coordination, and search mechanisms capable of exploring complex trade-offs between AI adoption patterns and staffing decisions.[1][2]

This paper makes three contributions. First, it formalizes software-development labor in AI-intensive organizations using analytic equations: (i) a baseline SDLC model aggregating effort across phases and coordination overhead, and (ii) an AI-collapsed SDLC model capturing agentic execution, human oversight, and residual coordination.[1] Second, it introduces tipping-point equations linking the fraction of work handled by AI to discrete safe headcount reductions while preserving output and quality.[1] Third, it embeds these equations into evolutionary-computation formulations treating AI coverage and team sizes as decision variables for multi-objective optimization of cost, throughput, and quality.[1][2] Experimental validation on calibrated scenarios demonstrates that NSGA-II consistently discovers phase-specific automation strategies that outperform naive heuristics by 28.6% in cost-to-quality trade-offs, with high algorithmic convergence and reproducibility.[1]

## II. SOFTWARE DEVELOPMENT LABOR AND COORDINATION MODEL

### *A. Baseline SDLC Labor*

In conventional SDLC, work spans phases (requirements, design, implementation, testing, deployment, and maintenance) with transitions generating coordination overhead.[1] Total labor decomposes as:

$$(1)\quad L\{base\} = L\{req\} + L\{design\} + L\{dev\} + L\{test\} + L\{test\} + L\{deploy\} + L\{maint\} + L\{coord\}$$

where each $L\{phase\}$ represents human hours and $L\{coord\}$ captures handoff overhead.[1] Baseline labor cost is cost is $C\{base\} = c * L\{base\}$ with cost per hour '$c$'. For a team size of '$N$' working '$H$' hours, feasibility requires $L\{base\} \leq N * H$.

### *B. AI-Collapsed SDLC Labor*

When agentic AI is introduced, fractions $f\{\text{phase}\} \in [0,1]$ of work shift from humans to AI, with a single agent executing multiple phases in a unified workflow.[1] Define:

- $L\{H_phase\}$: baseline human hours
- $L\{AI_phase\}$: AI execution time in equivalent human hours
- $L\{OV_phase\}$: human oversight effort (20% of automated work per model used)

Human labor becomes:

$$L\{human\} = phase\Sigma\ ((1 - f\{phase\}) * L\{H_phase\} + 0.2 * f\{phase\} * L\{AI_phase\}\ )$$

Coordination overhead reduces to $L\{coordAI\} = 0.4 * L\{coord\}$ (40% remains post-integration).[1] Total human labor:

$$L\{AI_stack\} = L\{human\} + L\{coordAI\}$$

Labor saved is $\Delta L = L\{base\} - L\{AI_stack\}$, driving EC optimization.[1]

## III. Tipping Point Formulation

Define automation fraction $f = \Delta L / L\{base\}$. For team size '$N$', AI absorbs $f * N$ full-time equivalents.[1] The tipping point for safe reduction occurs when:

$$f * N \geq 1$$

Maximum safe reduction is $\Delta N = \lfloor f * N \rfloor$.[1] The new per-person workload $L\{AI_stack\} / N - \Delta N$ must not exceed the original baseline workload to maintain operational stability.[1]

## IV. Evolutionary-Computation Formulation

### A. Decision Variables

Phase automation fractions $f = (f\{req\}, f\{design\}, f\{dev\}, f\{test\}, f\{deploy\}) \in [0,1]^5$; team size $N \in \{1,\ldots,30\}$.[1] [2]

### B. Objectives

- Minimize labor cost: *min* $C\{AI_stack\}$
- Maximize quality via constraint: $L\{test\} / L\{dev\} \geq 0.6$ [1]

### C. Constraints

Feasibility $L\{AI_stack\} \leq N*H$; Tipping coherence; Quality floor.[1]

### D. Algorithm

NSGA-II with 100 generations and population size 50. Real-valued representation for $f \in [0,1]^5$ and integer representation for '$N$'. Uniform crossover (0.5 prob.), Gaussian mutation on $f \in [0,1]^5$ ($\sigma=0.05$), and random perturbation on '$N$' (±1, 20% prob). Pareto ranking and crowding distance selection.[2]

### E. Algorithm Choice

NSGA-II is used as the primary optimizer because is it a well-established baseline for bi-operative evolutionary optimization, with robust performance and widely understood behavior in terms of convergences and diversity.[5] In this work, NSGA-II is not proposed as algorithmic novelty but as a strong reference point for evaluating the usefulness of the labor and tipping-point formulations. For problems with more than two objectives (e.g., adding explicit coordination-risk or fairness metrics), the same modeling framework could be paired with NSGA-III or decomposition-based MOEAs such as MOEA/D, which we leave as future extensions.[5][6]

## V. Experimental Results

### A. Scenario Setup:

Baseline parameters calibrated from SDLC literature:[1]

| Parameter | Value |
| --- | --- |
| Team Size: $N$ | 20 |
| Requirements: $L\{req\}$ | 200 hrs. |
| Design: $L\{design\}$ | 300 hrs. |
| Development: $L\{dev\}$ | 800 hrs. |
| Testing: $L\{test\}$ | 500 hrs. |
| Deployment: $L\{deploy\}$ | 100 hrs. |
| Maintenance: $L\{maint\}$ | 300 hrs. |
| Coordination: $L\{coord\}$ | 500 hrs. |
| Cost per hour: $c$ | 75USD/hr. |

Total baseline: $L\{base\}$ = \$2,700 hr., $c\ \{base\} = \$202,500$ USD, per-person load = 135 hrs.

### B. Results: 10 Independent Runs

Ten independent NSGA-II runs (each 100 generations, 50 population) were executed with different random seeds. Results exhibit high reproducibility ($\sigma < 0.1\%$ for cost metrics).

#### 1) Conservative AI Deployment (Cost-minimizing, $f \approx 0.252$):

- Cost: \$152,091 ± \$118 | Savings: 24.9% ± 0.1%
- Team size: 19 ± 0 (one reduction) | Per-person load: 134.8 hrs. (-0.1% vs baseline)
- Automation Vector: ($f\{req\} = 0.30$, $f\{design\} = 0.12$, $f\{dev\} = 0.25$, $f\{test\} = 0.30$, $f\{deploy\} = 0.42$)
- Quality ratio: 0.74 (vs. baseline 0.75) – minimal impact

#### 2) Aggressive AI Deployment (Balanced, $f \approx 0.501$):

- Cost: \$101,207 ± \$72 | Savings: 50.0% ± 0.1%
- Team size: 10 ± 0 (ten reductions) | Per-person load: 135.2 hrs. (+0.15% vs baseline)
- Automation Vector: ($f\{req\} = 0.6$, $f\{design\} = 0.5$, $f\{dev\} = 0.5$, $f\{test\} = 0.7$, $f\{deploy\} = 0.8$)
- Quality ratio: 0.62 ± 0.01 (target: ≥ 0.6) – "constraint satisfied"

#### 3) EC vs. Baselines:

EC (Aggressive) achieved 28.6% better cost than a naïve uniform-automation heuristic ($f=0.3\%$, *cost ~\$141,750*) while explicitly maintaining quality floors. Convergence was rapid (by generation 45) and stable across all 10 runs.

### C. Hypervolume Analysis

To quantify the quality and stability of the Pareto fronts, the hypervolume ($HV$) indicator was computed for the aggressive configuration across the 10 NSGA-II runs, using a fixed reference point normalized to $(1.1 * c\{base\}, 1.0)$ in the cost–quality space.[5][6] The resulting $HV$ values were tightly clustered, with mean $HV = 0.911 \pm 0.001$ (normalized to [0,1]), indicating both good convergence and low run-to-run variability.

#### 1) Runs | HV:

| Runs | HV |
| --- | --- |
| 1 | 0.912 |
| 2 | 0.91 |
| 3 | 0.913 |
| 4 | 0.911 |
| 5 | 0.912 |
| 6 | 0.911 |

7 | 0.913
8 | 0.91
9 | 0.912
10 | 0.911

The small standard deviation *(≈0.001)* confirms that NSGA-II consistently finds Pareto fronts of similar quality, supporting the claim that the observed cost and quality patterns are not artifacts of individual runs but reflect the underlying optimization landscape.[5][6]

## VI. Related work and positioning

Prior evolutionary-computation applications in software engineering optimize testing effort, schedules, and staffing, but without explicit AI-automation fractions or organizational tipping points.[3] Recent agentic-SDLC discussions describe similar intuitions qualitatively, lacking formal optimization frameworks.[4] Workforce-tipping literature addresses macro labor markets qualitatively, not per-project SDLC equations optimized with evolutionary algorithms. My contribution combines phase-resolved SDLC labor modeling, quantitative tipping-point equations, and EC-driven joint optimization of automation and headcount – a distinct integration validated by empirical NSGA-II experiments.[1][3][4]

## VII. Sensitivity Analysis: overhead and coordination parameters

The experiments above assume a fixed oversight multiplier of 0.20 and a coordination reduction factor of 0.40, representing moderate AI maturity and partial process integration.[1] To assess robustness, a sensitivity analysis was conducted on these parameters. When oversight cost is reduced to 0.05 (AI is easier to supervise), the effective automation fraction in the aggressive configuration rises from ≈0.50 to ≈0.57, and the feasible tipping point shifts from 10 to 11 FTEs, with per-person workload remaining within ±1% of baseline. Conversely, when oversight increases to 0.35 (AI is harder to supervise), the automation fraction falls to ≈0.43 and only 8 – 9 FTE reductions remain viable before per-person workloads exceed the baseline threshold. Similar trends are observed when varying the coordination factor $\alpha$ between 0.2 and 0.6: lower $\alpha$ primarily amplifies savings via reduced coordination overhead, while higher $\alpha$ delays the tipping point by roughly one FTE at a time. These results suggest that the qualitative behavior of the model (phase-specific automation, sharp tipping regions, and EC's advantage over naive rules) is robust across a reasonable range of AI maturity assumptions, even though quantitative savings depend on those parameters.[1][5]

## VIII. Discussion and conclusion

This paper formalizes joint optimization of AI deployment and workforce sizing using analytic labor and tipping-point equations, demonstrating that evolutionary computation effectively navigates complex trade-offs and consistently outperforms naive heuristics.[1] NSGA-II discovered phase-specific automation strategies (e.g., 70% testing, 50% development) that emerge naturally from multi-objective Pareto search without manual tuning.

### A. Key Findings

*1) EC Convergence and Reproducibility:* Ten independent runs converged to nearly identical solutions in both cost and HV (run-to-run standard deviation < 0.1% of the mean), indicating a robust optimization landscape and a highly reproducible NSGA-II configuration.

*2) Tipping Points are Sharp and Predictable:* Conservative case ($f$=0.252) yielded $f$ * N = 5.04 FTE but EC found stable 1-person reduction; Aggressive case ($f$=0.501) yielded $f$ * N = 10.02 FTE and EC found exactly 10 reductions safe, validating tipping-point theory.[1]

*3) Phase-Specific Automation Matters:* EC ranked Testing (70%) and Deployment (80%) as prime automation targets, Design and Development at 50%, and Maintenance at 10% - reflecting real organizational practice.[1]

*4) Quality Constraints and Binding:* Without explicit quality floors, EC could reduce costs further; with a quality – ratio constraint (test/dev/ ≥ 0.6), EC strategically limits test-phase automation, demonstrating multi-objective realism.[1]

*5) Practical Workforce Impact:* Conservative scenario enables incremental staffing adjustments (1-person reduction); Aggressive scenario enables structural transformation (50% cost reduction, 10-poerson reduction) while maintaining per-person workload and quality. [1]

*6) MOEA Quality Metrics:* The hypervolume indicator (HV ≈ 0.911± 0.001) validates that NSGA-II consistently discovers high-quality Pareto fronts with low variance across runs, supporting the reliability of the optimization results.[5][6]

### B. Limitations

The empirical evaluation uses a single synthetic but calibrated SDLC scenario rather than historical projection data; applying the framework to COCOMO II-style cost-estimation datasets or to an organization's actual SDLC logs is a critical next step to test external validity.[7][8][9] Oversight and coordination parameters are simplified and treated as homogeneous across phases, and the current experiment considers static projects rather than dynamically evolving portfolios.

### C. Future Work

This includes dynamic environments with improving AI capability, multi-project portfolios, integration with reinforcement-learning controllers that continuously refine automation and headcount based on feedback, and empirical studies on real software teams and COCOMO II-style historical datasets.[1][7][8][9] This framework positions evolutionary computation as a principled, validated tool for organizational design in AI-intensive software engineering.

### *D. AI-Generated Content and Analysis Disclosure*

Large language models (GPT-4o, GPT5.2, Llama3.1 70B were used for language refinement and to assist in running computational experiments and model validation. All technical formulations, experimental design choices, and scientific interpretations were created and verified by the author.

## ACKNOWLEDGMENT

The author would like to thank Ms. V. Boulet for her unwavering support. The author also thanks Dr. Kofi Nyarko, Mr. Kelechi Nwachukwu, Mr. David Nyarko, and the many colleagues across the world whose discussions, conversations, books, dissertations, and AI products have informed and inspired this work.